\documentclass[fleqn,10pt]{wlscirep}

\title{On the relation between UHI intensity and city proximity}

\usepackage{bigints}

\author[1]{Bin Zhou}
\author[1,*]{Diego Rybski}
\author[1,2]{J{\"u}rgen P. Kropp}
\affil[1]{Potsdam Institute for Climate Impact Research, Potsdam, Germany}
\affil[2]{Department of Geo- and Environmental Sciences, University of Potsdam, Potsdam, Germany}

\affil[*]{ca-dr@rybski.de}

\begin{abstract}
Recently, D.~Zhou~et~al.~2015\cite{ZhouZZSL2015} studied empirically the land surface temperature of 32 Chinese cities and reported an \emph{exponentially decaying} residual effect of the urban heat island (UHI) around the cities. 
Here we show analytically that such a form is equivalent to a previously proposed two-dimensional \emph{Gaussian function} \cite{StreutkerDR2002,StreutkerDR2003}.
The reason for this seeming contradiction is the way how the distance from the considered city is defined.
While in the former, consecutive equal area belts around the city are used, in the latter it is the euclidean distance.
In simplified terms, the definition of the belts implies a transformation of the independent variable $\rho\sim d^{\alpha}$ with $\alpha\approx 1/2$, where $\rho$ is the euclidean distance from the center and $d$ is the index of the belt.
Since the belts have equal area, outer ones become more narrow.
As a consequence, the Gaussian function $\sim \mathrm{e}^{-\rho^2}$ becomes an exponential cone $\sim \mathrm{e}^{-d}$.
This paper provides an explicit derivation of the equivalence.
\end{abstract}

\begin{document}

\flushbottom
\maketitle

\thispagestyle{empty}

\section*{Introduction}

The urban heat island (UHI) effect is a well studied phenomenon and represents an illustrative example of human-environment-interaction\cite{Oke1987,Arnfield2003,VoogtO2003}. 
At the interface between built-up and natural land-cover a temperature gradient emerges that in many cases is decaying away from the city.
Despite being known for a long time and being investigated for decades, estimates of the UHI intensity are sensitive to the choice of indicators\cite{SchwarzLS2011}.
Crucial to the quantification are (i) the employed city definition and (ii) the overlying temperature field.

Here we review two approaches characterizing the (surface) temperature field in the vicinity of cities and unify them. 
Our work consists of an analytical derivation based on previous empirical studies.
Nevertheless, our result contributes to a better understanding of the UHI phenomenon and could support a standardized UHI assessment.

\section*{Results}

D.~Zhou~et~al.~2015\cite{ZhouZZSL2015} defined the cities by means of polygons of large fraction of built-up areas as obtained from Landsat data. 
Around each polygon consecutive (non-cummulative) belts were defined, whereas their areas are half the area of the city.
Next, using MODIS land surface temperature data, averages were calculated within each belt.
Similar to previous studies\cite{PengPCFOBNZM2012,ZhouRK2013}, the UHI intensity is then calculated as the difference between the belt temperature and a reference temperature, i.e.\ $\Delta T= T_\text{b}(d)-T_\text{ref}$, where $T_\text{b}(d)$ is the average temperature in the belt with index~$d$ and the $T_\text{ref}$ is the average over the 3 outermost belts.
Plotting the UHI intensity $\Delta T$ versus the belt index $d$ an exponential decay according to $\Delta T = A\,\mathrm{e}^{-S\,d}+T_0$
was found\cite{ZhouZZSL2015}, where $A$ is the maximum temperature difference at the city boundary, $S$ is the decay rate, and $T_0$ is an offset.
The authors consider $S$ as UHI footprint. 

Since both, $T_\text{ref}$ and $T_0$, are constant, we consider the exponential temperature decay
\begin{equation}
T_\text{b}(d)= A\,\mathrm{e}^{-S\,d}
\label{eq:expdecay}
\end{equation}
outside the city with increasing belt index, and for simplicity we disregard any offset.

\begin{figure}[ht]
\centering
\includegraphics[width=0.65\linewidth]{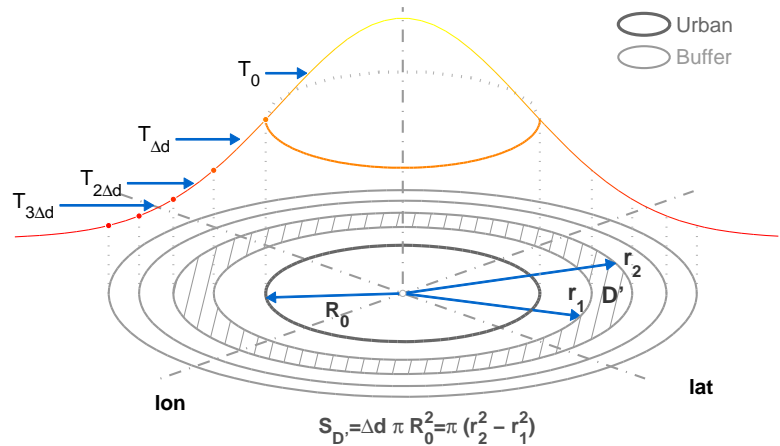}
\caption{Illustration of the two-dimensional Gaussian function and the belts around the city. Here only the case $r>R_0$ is considered.}
\label{fig:illu}
\end{figure}

In order to prove that Eq.~(\ref{eq:expdecay}) is equivalent to a Gaussian function, we approximate the city as a circle with radius $R_0$ and ring-shaped belts extending between the radii $r_1$ and $r_2$ (see Fig.~\ref{fig:illu}), whereas according to D.~Zhou~et~al.~2015\cite{ZhouZZSL2015} the area of the belts is a multiple of the city area, i.e.\ 
\begin{equation}
\Delta d=\frac{\pi r_2^2-\pi r_1^2}{\pi R_o^2}
\end{equation}
with chosen\cite{ZhouZZSL2015} $\Delta d=0.5$ which is kept fixed.
Because of this ``equal area definition'', belts close to the city are wider and those far away become narrow.

We start from the Gaussian function which in polar coordinates is given by
\begin{equation}
g(\rho,\phi) = B\mathrm{e}^{-\frac{\rho^2}{2\sigma^2}}
\label{eq:2dgau}
\, ,
\end{equation}
where $\rho$ and $\phi$ are radius and angle, respectively, and $B$ and $\sigma$ are the amplitude, and spatial extent, respectively.
The mean temperature over an area $D$ is given by
\begin{align}
T_D &=\frac{\iint_D g(\rho,\phi)\rho\,\mathrm{d}\rho\mathrm{d}\phi}{\iint_D \rho\,\mathrm{d}\rho\mathrm{d}\phi } \nonumber \\
	&= \frac{ \int_0^{2\pi}\int_{0}^{R_0} B\mathrm{e}^{-\frac{{\rho}^2}{2\sigma^2}}\rho\,\mathrm{d}\rho\mathrm{d}\phi}{\int_0^{2\pi}\int_{0}^{R_0}\rho\,\mathrm{d}\rho\mathrm{d}\phi} \nonumber
\end{align}

If $d=0$ denotes the urban area, the belt zone with  $d =\{\Delta d, 2\Delta d, \dots \}$ away from the urban outline is $D'= \{(\rho, \phi) \in \mathbb{R^+}^2: r_1 \leqslant \rho\leqslant r_2,0 \leqslant \phi \leqslant 2\pi, r_2^2-r_1^2=\Delta dR_0^2, r_2^2=(d+1)R_0^2 \}$.
Making the substitution $u = - \frac{\rho^2}{2\sigma^2}$ leads to $\mathrm{d}\rho =-\frac{\sigma^2}{\rho}\mathrm{d}u$.
Then the mean temperature of the belt zone $D'$ is given by
 \begin{align}
 T_{D'} &=\frac{-B\sigma^2\bigint_{-\frac{r_1^2}{2\sigma^2}}^{-\frac{r_2^2}{2\sigma^2}}\mathrm{e}^u\,\mathrm{d}u\bigint_{0}^{2\pi}\mathrm{d}\phi}{\pi (r_2^2-r_1^2)} \nonumber\\
 &=\frac{2B\sigma^2(\mathrm{e}^{-\frac{r1^2}{\sigma^2}}-\mathrm{e}^{-\frac{r_2^2}{2\sigma^2}})}{r_2^2-r_1^2} \nonumber \\
  &=\frac{2B\sigma^2(\mathrm{e}^{-\frac{r1^2}{2\sigma^2}}-\mathrm{e}^{-\frac{r_2^2}{2\sigma^2}})}{r_2^2-r_1^2} \cdot \frac{\mathrm{e}^{\frac{r_2^2}{2\sigma^2}}}{\mathrm{e}^{\frac{r_2^2}{2\sigma^2}}}\nonumber\\
  &=\frac{2B\sigma^2(\mathrm{e}^{\frac{\Delta d\,R_0^2}{2\sigma^2}}-1)}{\Delta d\,R_0^2\,\mathrm{e}^{\frac{(d+1)R_0^2}{2\sigma^2}}} \nonumber\\
  &=\frac{2B\sigma^2(\mathrm{e}^{\frac{\Delta d\,R_0^2}{2\sigma^2}}-1)}{\Delta d\, R_0^2\,\mathrm{e}^{\frac{R_0^2}{2\sigma^2}}} \mathrm{e}^{-\frac{R_0^2}{2\sigma^2}d} \nonumber \\
  &= B'\,\mathrm{e}^{-\frac{R_0^2}{2\sigma^2}d} \label{eq:proof} \\
  & \nonumber \hspace{15cm} \square
 \end{align}
As $\Delta d$, $B$, $R_0$ are predefined or city-specific constants, also $B'={(2B\sigma^2(\mathrm{e}^{\frac{\Delta d\,R_0^2}{2\sigma^2}}-1))}/{(\Delta d\, R_0^2\,\mathrm{e}^{\frac{R_0^2}{2\sigma^2}})}$ is constant, so that Eq.~(\ref{eq:proof}) resembles Eq.~(\ref{eq:expdecay}).
Thus, we conclude that the exponential decay Eq.~(\ref{eq:expdecay}) is equivalent to the Gaussian function Eq.~(\ref{eq:2dgau}).
This equivalence can be derived analogously starting from the exponential decay, Eq.~(\ref{eq:expdecay}).

\section*{Discussion}
In summary, we have proven analytically that the exponential decay with belt index\cite{ZhouZZSL2015} is approximately equivalent to the Gaussian function with euclidean distance\cite{StreutkerDR2002,StreutkerDR2003}.
It is important to mention that both approaches were proposed for land surface temperature. 
It cannot be excluded that the empirical quantification and resulting parameters will be different for 2m air temperature.
Moreover, due to noise in the data, showing the validity of any of both shapes requires a thorough statistical analysis. 
Here we only prove the theoretical approximate equivalence of both descriptions.
An important assumption made is that the cities are approximately circular. 
Many cities also exhibit fractality and anisometry\cite{ZhouRK2017}, however, for the outermost belts the assumption of circularity is better justified.

\section*{Methods}
In this paper we use solely analytical derivation.

\section*{Acknowledgments}

The research leading to these results has received funding from the European Community's Seventh Framework Programme under Grant Agreement 308497 (Project RAMSES). 
The authors thank anonymous reviewers for helpful comments on the manuscript.
Author BZ thanks Climate-KIC, the climate innovation initiative of the EU's European Institute of Innovation and Technology (EIT), for award of a Ph.D.\ scholarship.

\end{document}